\documentstyle{pss}
\input epsf.tex

\newcommand{\eg}{\mbox{e.\ g.\ }}

\newcommand{\be}{\begin{eqnarray}}
\newcommand{\en}{\end{eqnarray}}

\newcommand{\mc}{\mathcal}

\newcommand{\f}{f}  
\newcommand{\p}{p}
\newcommand{\fd}{f^{\dagger}}
\newcommand{\pd}{p^{\dagger}}
\newcommand{\at}{\tilde{a}}
\newcommand{\atd}{\tilde{a}^{\dagger}}

\begin{document}

\title{Quantum and thermal effects in the double 
exchange ferromagnet}

\author{Nic Shannon}

\address{Max--Planck--Institut f{\"u}r Physik komplexer Systeme,
N{\"o}thnitzer Str. 38, 01187 Dresden, Germany}

\submitted{ --- Physica Status Solidi (b) {\bf 236}, 494 (2003)} 
\maketitle

\hspace{9mm} Subject classification: 75.25.+z
              
\begin{abstract}
The physics of the ferromagnetic phase of the 
``double exchange'' model has been widely discussed 
in the context of the CMR manganites.  Usually, 
the double exchange ferromagnet is treated is
classically, by mapping it onto an effective 
Heisenberg model.
However this mapping does not permit a 
correct treatment of quantum or thermal fluctuation
effects, and the results obtained lack
many of the interesting features seen in 
experiments on the manganites.
Here we outline a new analytic 
approach to systematically evaluating quantum and
thermal corrections to the magnetic and electronic 
properties of the double exchange ferromagnet.
\end{abstract}

\section{Ferromagnetic manganites and double exchange}

Systems in which charge carriers move in a background of 
local magnetic moments are ubiquitous among doped d--electron 
oxides, the two most widely studied examples being the colossal 
magnetoresistance (CMR) manganites and the high temperature
superconducting (HTc) cuprates.

It was first suggested by Zener \cite{zener}
that ferromagnetism in the 
manganites originated from the kinetic energy of the doped 
holes, which couple to the local moments of the 
Mn atoms through a strong Hund's first rule coupling.
This idea was later formalised
in terms of the strong coupling limit of a ferromagnetic 
Kondo lattice Hamiltonian \cite{anderson,degennes}
\be
\label{eqn:KondoH}
{\mc H}
  = -t \sum_{\langle ij \rangle\alpha} 
           c^{\dagger}_{i\alpha} c_{j\alpha}
       -J \sum_{i} \vec{S}_i.\vec{s}_i 
   &\qquad& 
\vec{s}_i = \frac{1}{2}\sum_{\alpha\beta} 
      c^{\dagger}_{i\alpha}
      \vec{\sigma_{\alpha\beta}}c_{i\beta}
\en
Here, $t$ is the hopping integral 
for $e_g$ electrons transfered between neighbouring 
Mn atoms, whose half filled $t_{2g}$ levels are modelled
as a spin $S=3/2$ local moment, and $J \gg t > 0$ is the 
Hund's rule coupling between itinerant and localised 
electrons.
The orbital degeneracy of the $e_g$ levels (which are
Jahn--Teller active) has been neglected, 
as have next nearest neighbour hoppings and all
Heisenberg super--exchange processes.  This minimal
model --- a single band of tight binding electrons
coupled to a background of local moments is for historical
reasons known as the ``double exchange'' (DE) model, and
we will refer to its Ferromagnetic ground state as the 
``double exchange ferromagnet'' (DEFM).

The model Equation (\ref{eqn:KondoH}) has been widely studied for 
fifty years, and we shall not pretend here to offer a review
of either its diverse physical properties, or of the many
ingenious approaches which have been used to study them.
Rather, we present a concise introduction to a simple
and controlled way of performing a spin wave expansion 
for such a Hamiltonian, and a brief overview of the effect
of quantum and thermal fluctuations on the DEFM.
We believe that the methods developed can easily be transferred
to other magnetic d--electron oxides. 

\section{Spin wave Expansion}

In the strong Hund's rule coupling limit $t/J \to 0$,
so far as magnetic properties are concerned, it makes little
sense to distinguish between localised and itinerant electrons.
For example, oxidised Mn atoms in the canonical 
CMR manganite La$_{1-x}$Ca$_x$MnO$_3$ can exist either as 
Mn$^{4+}$ of Mn$^{3+}$ ions, which are respectively spin 
S=3/2 and spin S=2 local moments.  Where the Hund's rule coupling
is so strong, it is therefore convenient to work in the local
basis of eigenstates of total (i.e. ``itinerant'' $e_g$ plus 
``localised'' $t_{2g}$ electron) spin $\vec{T} = \vec{S} + \vec{s}$.  

Taking this as a starting point, we have derived a controlled
large $S$ expansion of Equation~(\ref{eqn:KondoH}) which 
can be applied to any magnetically ordered state.  
Our original approach \cite{EPL,shannon,PRB,salerno} was to generalise
the Holstein--Primakoff (HP) transformation \cite{holstein} by
allowing the length of the fluctuating spin T to be an 
operator quantity, and introducing local ``up''($\f$) and 
local ``down'' ($\p$) electron operators such that 
$T = S + (\fd\f - \pd\p)/2$.  
A more general derivation of the associated spin and charge
algebra based on the Schwinger Boson representation of Penc and 
Lacaze \cite{karlo1} will be presented elsewhere \cite{karlo2}.  
 
\begin{figure}[tb]
\begin{center}
\leavevmode
\epsfysize 4cm
\epsffile{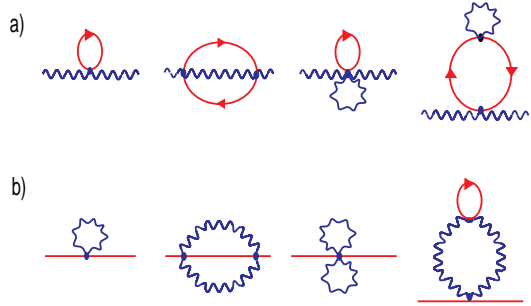}
\caption{
Diagrams contributing to the self energy of a) spin waves
and b) low energy spin--projected electrons 
to order ${\mc O}(1/S^2)$.}
\label{fig:watermelon}
\end{center}
\end{figure}

Where the operator $\fd$ acts on an empty site, it creates 
a low energy state with total spin $T=S+1/2$, while the 
the operator $\pd$ creates a high energy state with total spin 
$T=S-1/2$.
All information about low energy spin excitations is encoded
in the generalised Holstein--Primakoff bosons $\at$, so in the 
limit $t/J \to 0$, we can project out the high energy, 
low spin states associated with $\p$ electrons altogether, 
and obtain a very simple representation 
of Equation (\ref{eqn:KondoH}) :
\be
\label{eqn:effectiveH}
{\mc H}^{\prime} &=&  \sum_{k_1} \epsilon_1 \fd_1 \f_1
       +  \frac{1}{N} \sum_{k_1\ldots k_4}  v^{13}_{24}
 \fd_1 \f_2 \atd_3 \at_4 \delta_{1+3-2-4} \\
&& v^{13}_{24} =  \frac{1}{4(S+\frac{1}{2})}
   \left[ 
   \left( 1 + \frac{1}{8S} \right) 
   \left( \epsilon_{1+3} + \epsilon_{2+4}\right)
   - \left(\epsilon_1 + \epsilon_2\right)
   \right]
\en
where $\epsilon_k$ is the (tight binding) electron dispersion 
and $v^{13}_{24}$ the vertex for interaction between band
electron ($\f$) and spin wave modes ($\at$). 
For compactness, a further four--boson/two--fermion vertex 
has been dropped.  The generalisation of this Hamiltonian
to bilayer systems is straightforward \cite{EPJB}.

From this starting point it is possible to
use a conventional diagrammatic perturbation theory to evaluate 
spin--wave and electron self energy corrections.  The leading
zero temperature semi--classical process is the one loop
diagram of Figure (\ref{fig:watermelon}a), which generates a 
dispersion for the spin waves, and so stabilises
the FM ground state.  

\section{Quantum and Thermal Effects}

At a semi--classical level (i.e. considering only the ${\mc O}(1/S)$
one loop diagrams of 
Figure (\ref{fig:watermelon}a and \ref{fig:watermelon}b))  
the DEFM can be mapped onto an independent set of spin excitations,
described by a nearest--neighbour Heisenberg model 
(see \eg \cite{degennes,PRB}), and a band of tight binding 
electrons with temperature dependant bandwidth \cite{nic}.

\begin{figure}[h]
\epsfysize 5cm
\epsffile{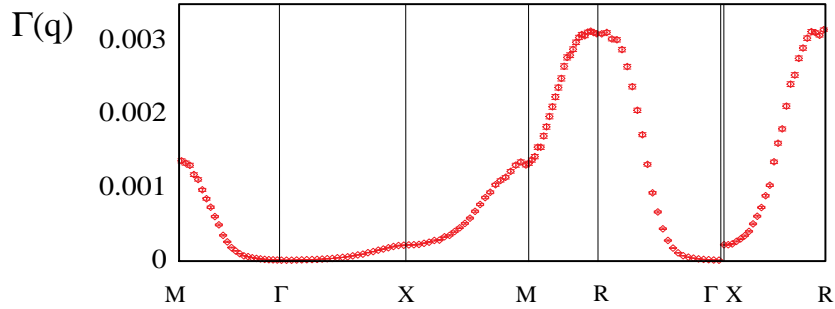}
\caption{Damping of spin waves in a DEFM with electron doping 
x=0.7, at zero temperature, throughout the Brillouin zone.  
Energies are measured in units of the electron bandwidth 
$2zt=1.0$.}
\label{fig9}
\end{figure}

\begin{figure}[h]
\epsfysize 5cm
\epsffile{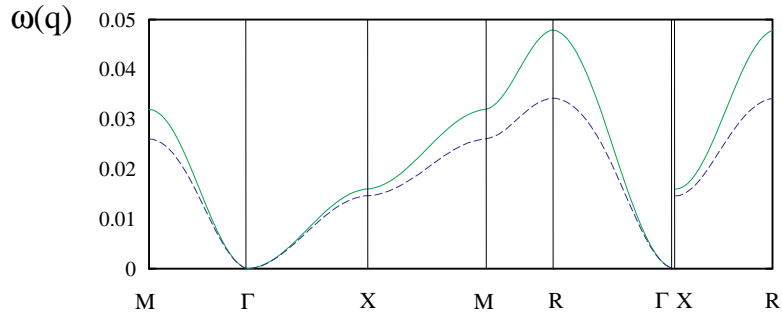}
\caption{Spin wave spectrum of the DEFM with electron doping 
x=0.7 at zero temperature.  
Upper line --- semi--classical result.
Lower line --- result when leading quantum corrections are taken 
into account.  
Energies are measured in units of the electron 
bandwidth $2zt=1.0$.
}
\label{fig8}
\end{figure}

Once the leading quantum effects are taken into account
(the remaining ${\mc O}(1/S^2)$ diagrams of 
Figure (\ref{fig:watermelon})), this mapping breaks down.
The spin waves become damped (see Figure (\ref{fig9})), 
even at zero temperature,
and both quantum and thermal fluctuations 
dynamically generate new non--nearest neighbour couplings
between spins, which progressively modify the form of
spin wave dispersion (see Figure (\ref{fig8})).

\begin{figure}[h]
\epsfysize 5cm
\epsffile{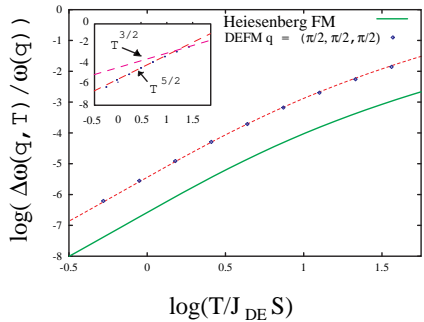}
\caption{Log--log plot of finite temperature corrections to 
the spin wave spectrum of the DEFM, together with those
of the semi--classically equivalent Heisenberg FM.  
Temperature corrections
$\Delta\omega(q,T)$ are normalised to bare dispersion $\omega(q)$,
temperatures to the scale of spin wave dispersion $J_{DE}S$.
}
\label{fig10}
\end{figure}

Thermal corrections to the overall scale of the spin wave
dispersion in the DEFM are broadly similar to those in 
the Heisenberg FM, but are considerably enhanced relative
to the classically equivalent Heisenberg model (see Figure 
(\ref{fig10})).

At zero temperature, in the limit $t/J \to 0$, the low energy 
$\f$ electron states remain undamped.  However at finite 
temperature they interact through the transverse spin 
susceptibility, and have a finite lifetime which is controlled 
by the phase space for spin fluctuations \cite{nic}.

\section{Conclusions}

We have shown that it is possible to derive a controlled large $S$ 
expansion for the Kondo lattice which, when applied to the DEFM
in the limit $t/J \to 0$, reduces the model to a set of
composite spin modes interacting with spinless electrons.  At
a semi--classical level spin and charge modes do not interact,
but at ${\mc O}(1/S^2)$ interactions are restored and the 
physics of the system substantially modified.

At a qualitative level, the departures from Heisenberg model
behaviour caused by quantum and thermal fluctuation effects
in the DEFM resemble those seen in inelastic neutron scattering
experiments on the CMR manganites.  However there remains some
doubt that a minimal model of the form Equation (\ref{eqn:KondoH})
is sufficient to fit experimental data for the manganites 
quantitatively \cite{EPJB}.

\end{document}